\documentclass[aps,prl,groupedaddress]{revtex4}%,twocolumn
\usepackage{revsymb}
\usepackage[latin1]{inputenc}
\usepackage[T1]{fontenc}
\usepackage[centertags]{amsmath}
\usepackage{amstext,amssymb}
\usepackage{graphicx}
\usepackage[footnotesize,center]{caption2}
\usepackage{units}
\usepackage{fancyhdr}
\usepackage{nonfloat}

\newcommand{\bild}[5]{\begin{figure}[#3]
                        \centering
                        \includegraphics*[width=#2]{#1}%,height=#2
                        \sl\caption{#4}
                        \label{#5}
                        \end{figure}}
\newcommand{\w}{\omega}

\usepackage{xspace}

\newcommand{\Figref}[1]{Fig.~\ref{#1}}
\newcommand{\eqnref}[1]{eq.~\ref{#1}}
\newcommand{\enquote}[1]{``#1''}

\begin{document}

\title{Arrayed Continuous-wave THz Photomixers}
\author{S. T. Bauerschmidt}\author{  G. H.
D\"ohler}
\affiliation{Max Planck Institute for the Science of Light, D-91058 Erlangen, Germany}
%\author{L. J. Wang}
%\affiliation{Department of Precision Instruments and Mechanology, Tsinghua University, Beijing, China}%Department of Precision Instruments and Mechanology
\author{H. Lu}\author{  A. C. Gossard}
\affiliation{Materials Department, University of California, Santa
Barbara, California 93106}
\author{S. Malzer}\author{  S. Preu$^*$}
\affiliation{Chair for Applied Physics, Friedrich-Alexander University Erlangen-Nuremberg, D-91058 Erlangen, Germany, *e-mail: Sascha.preu@physik.uni-erlangen.de}

\begin{abstract}
We present both chip-scale and free space coherent arrays of continuous-wave THz photomixers. By altering the relative phases of the exciting laser signals, the relative THz phase between the array elements can be tuned, allowing for beam steering.
The constructive interference of the emission of $N$ elements leads to an increase of the focal intensity by a factor of $N^2$ while reducing the beam width by $\sim N^{-1}$, below the diffraction limit of a single source. Such array architectures strongly improve the THz power distribution for stand-off spectroscopy and imaging systems while providing a huge bandwidth at the same time. We demonstrate this by beam profiles generated by a free space 2$\times$2 and a 4$\times$1 array for a transmission distance of 4.2 meters. Spectra between 70 GHz and 1.1 THz have been recorded with these arrays. 
\end{abstract}

%\ocis{220.4830 Systems design; 110.6795 Terahertz imaging; 130.3120 Integrated optics devices; 300.6495 Spectroscopy, terahertz }
\maketitle

Applications of Terahertz (THz, 100 GHz-10 THz) radiation in spectroscopy, security and imaging have gained strong interest in recent years due to the unique properties of matter and radiation in this spectral range. Many gases and substances show very narrow, unique spectral absorption features that allow for unambiguous identification \cite{Brown}. %Plusquellic07,ppm-Det2004,
 THz spectra even allow for identifying different isotopologues of the same chemical compound \cite{D2O}. %NeuesvonChristian,
Furthermore, THz radiation penetrates through common clothing and can reveal hidden objects underneath, even non-metallic threats \cite{ZhangSecurity,Explosives}. These properties are very attractive for security demands, particularly in terms of detection of explosives and drugs. Such applications, however, require both sensitive and fast detectors and powerful emitters. Different emitter concepts like frequency multiplied Gunn- or IMPATT-diodes \cite{Gunn1p7-1p9}, quantum cascade lasers \cite{QCL_lowf}, Josephson junction superlattices \cite{LutfiSupST} or backward wave oscillators \cite{BWOImaging04} have already been realized, but they are often limited in their tunability or require a cryogenic environment. For accessing a large frequency range, photomixing \cite{MeinReview,RubenEL,APL1,ITO_SST,Book_THzGeneral} is the method of choice, covering several octaves with a single setup at room temperature. In most cases, pulsed time domain systems \cite{GasMittleman98,Sartorius08,Schall_Helm_Keiding} are used since they provide very high output power levels \cite{DekorsyNeu,MeinLAE} with up to a few mW average THz power. Continuous-wave (CW) systems operating with a single device are yet limited to the $\mu$W range at 1 THz \cite{ITO_SST,RenaudSeedsHighP}.

A possible way to overcome the output power limitation of a single CW emitter utilizes a phased antenna array of individual, mutually coherent sources. Focal plane arrays \cite{focalPlaneArray} are already used in the THz and sub-THz frequency range \cite{mmWTHzimaging} for detection in both passive and active imaging. Phased emitter arrays are described in RF and microwave theory %\cite{Balanis} 
and are frequently used as directive elements to transmit electromagnetic radiation. In a similar manner, phase coherent THz sources can be set up in an array configuration and even be integrated on a chip \cite{ImagingArray,Integrated}.

In this paper, we show two versions of such an array: first, we show an on-chip array consisting of low-temperature grown GaAs (LT-GaAs) photoconductive mixers. Second, we show a free space array of individually packaged n-i-pn-i-p superlattice photomixers \cite{APL1,MeinReview}.  We discuss the advantages of both continuous-wave setups. All experiments were carried out at room temperature.

\bild{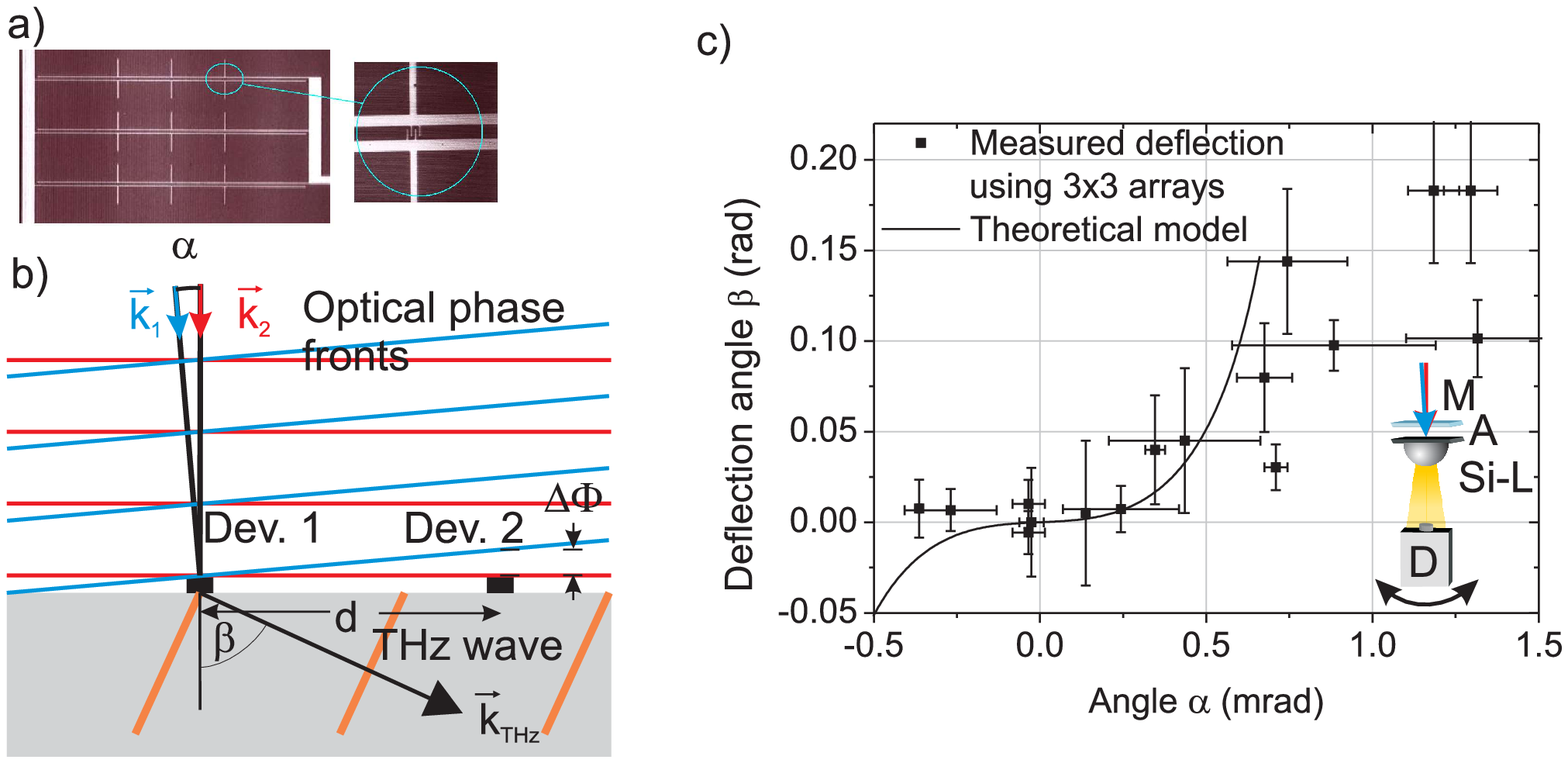}{12cm}{h}{a) On-chip-LT-GaAs array with dipole antennas. The dipoles of length 190 $\mu$m are spaced by 250 $\mu$m. b) Geometry, optical and THz phase fronts. A phase shift of $\Delta\phi$ is introduced between device 1 and device 2, resulting in off-axis emission of the array. c) THz beam defelection vs. inclination angle between the laser beams, $\alpha$, at 0.31 THz. The deflection angle saturates at 0.175 rad= 10$^\circ$ due to the focusing properties of the silicon lens attached to the array. The solid line is an approximative calculation of the beam deflection by ray-tracing. It is only valid for small deflection angles where it reproduces the experiment well. The inset shows the experimental setup: M: microlens array, A: dipole array, Si-L: silicon lens, D: Golay cell detector.}{Fig1}

The on-chip version is illustrated in \Figref{Fig1} a. It consists of LT-GaAs photoconductive switches with a finger electrode geometry (gap: 2$\mu$m, width 1$\mu$m, total area: 10$\times$10 $\mu$m$^2$) and resonant dipole antennas \cite{SPIE06}. The emitters are arrayed in a 3$\times$3 configuration. DC bias lines connect the emitters. Two Ti-sapphire lasers, at frequencies $\nu_1$ and $\nu_2=\nu_1+\nu_{THz}$ are heterodyned in order to provide a THz beat note at the difference frequency, $\nu_{THz}$. In order to adjust the THz phase between the emitters, we allow for a small angle, $\alpha$ between the two free space laser beams (see \Figref{Fig1} b). The total laser intensity becomes

\begin{equation}
I(x,t)=I_0\left[1+\cos\left(\Delta \vec k\cdot \vec x- \Delta \w t+\Delta
 \varphi\right)\right],\label{beat}
\end{equation}

where $\Delta\vec{k}\cdot \vec{x}=(\vec{k}_1-\vec{k}_2)\cdot \vec{x}$ is a spatially varying phase due to the inclination of the laser beams; $\Delta \w =2\pi\left(\nu_2- \nu_2 \right)=\w_{THz}$ is the (angular) THz frequency, and $k_{1,2}=2\pi\nu_{1,2}/c$ are the wave vectors of the optical laser beams.  $\Delta\varphi$ is an (arbitrary) overall phase between the two laser signals, which can be chosen zero. The relative THz phase of two photomixers that are separated by a distance, $d$, is

\begin{equation}
\Delta\Phi=\Delta\vec{ k}\cdot \vec d=(k_2\sin\alpha_2 -k_1 \sin\alpha_1)\cdot d, \label{phase_diff}
\end{equation}

where the anlges $\alpha_i$ refer to the angle between the laser beam and the surface normal. For simplicity, one beam incidence is chosen perpendicular to the device, i.e. $\alpha_2=0$ and $\alpha_1=\alpha$ is the inclination angle of the other laser beam as shown in \Figref{Fig1} b). This results in a phase difference of $\Delta\Phi=k_1d \sin\alpha=2\pi\sin\alpha\cdot (d/\lambda_1)$, where $\lambda_1$ is the laser wavelength ($\sim$ 850 nm). The THz phase is therefore \enquote{amplified} by the ratio of laser wavelength and the emitter spacing, $d\sim\lambda_{THz}$ , which is roughly a factor of 1000. Very small angles between the two heterodyned lasers are sufficient to achieve a large phase shift of the emitted THz beam. For example, an angle of $\alpha=0.2^\circ$ is sufficient to achieve a $2\pi$ phase shift for an emitter pitch of $d=$250 $\mu$m and a laser wavelength of 850 nm, as used in this experiment. For equidistant devices in the array, the single parameter $\alpha$ is sufficient to adjust all phases correctly for coherent emission in one direction. This allows for rapid scanning of the THz beam by a small angular offset between the optical beams. 
The challenge of on-chip arrays is the distribution of the optical power on the individual elements, since the active device area ($\sim 100$ $\mu$m$^2$) is only a very small fraction of the illuminated array area ($\sim 0.5$ mm$^2$). A microlens array, adapted to the pitch of $d=$250 $\mu$m is therefore used to distribute the heterodyned laser signals. For outcoupling of the THz beam, a hyperhemispherical silicon lens with a diameter of 10 mm was used. The lens focuses the THz beam due to the hyperhemisphericity. Both spherical aberrations and the focusing properties decrease the total deflection angle. \Figref{Fig1} c) shows the deflection of the THz beam vs. inclination angle between the optical beams, $\alpha$, for 3$\times$3 arrays with a pitch of $d=250$ $\mu$m, demonstrating the steerability of the THz beam.

%\bild{Fig2.eps}{13cm}{h}{[links Winkel in grad] }{Fig2}

%The Taylor expansion of $k_2$ with respect to the THz wavelength provides a phase difference of $\Delta\Phi=k_1d(\sin\alpha\lambda_{THz}/\lambda_1)$ [check], where $\lambda_1$ is the laser wavelength ($\sim$ 850 nm) and $\lambda_{THz}$ is the THz wavelength ($\sim$ 1 mm). 

On the one hand, integrating photomixers on a single chip bears the advantages of small size and the integratability with a silicon lens. For large arrays with a huge number of elements, however, a silicon lens may not be necessary at all, since a high directivity ensures efficient outcoupling through the backside of the chip \cite{QuiqueGo}. On the other hand, the distribution of the optical power on the very small sized photomixers (typically $\sim 10\times 10$ $\mu$m$^2$) requires careful alignment. Furthermore, cross-talk between the arrayed antennas is inevitable: Chip-integrated antenna elements are typically spaced on the same scale as the THz-wavelength. In addition, semiconductor based THz photomixer sources require electrical biasing that has to be integrated in the layout. The design of integrated arrays requires simulations. Optimum performance is typically only achievable in a narrow bandwidth and steerability is limited since phase shifts between devices affect the cross talk. Broadband operation with reduced cross talk would require larger spacing between the elements. If the array elements are too loosely spaced, the side lobes will gain strength \cite{QuiqueGo}.

Another possibility to gain directivity and brightness is an array of mutually coherent free space sources, each with relatively small aperture. \Figref{Fig2} shows a typical setup. There is no RF cross-talk since the emitters can be electrically isolated from each other. Both large bandwidth and high power are achieveable. Furthermore, the mutual coherence results in an interference pattern which can be optimized for small spot sizes at stand-off distances. Similar to the on-chip array, all photomixers are driven by the same pair of lasers. This assures the mutual coherence. 

\bild{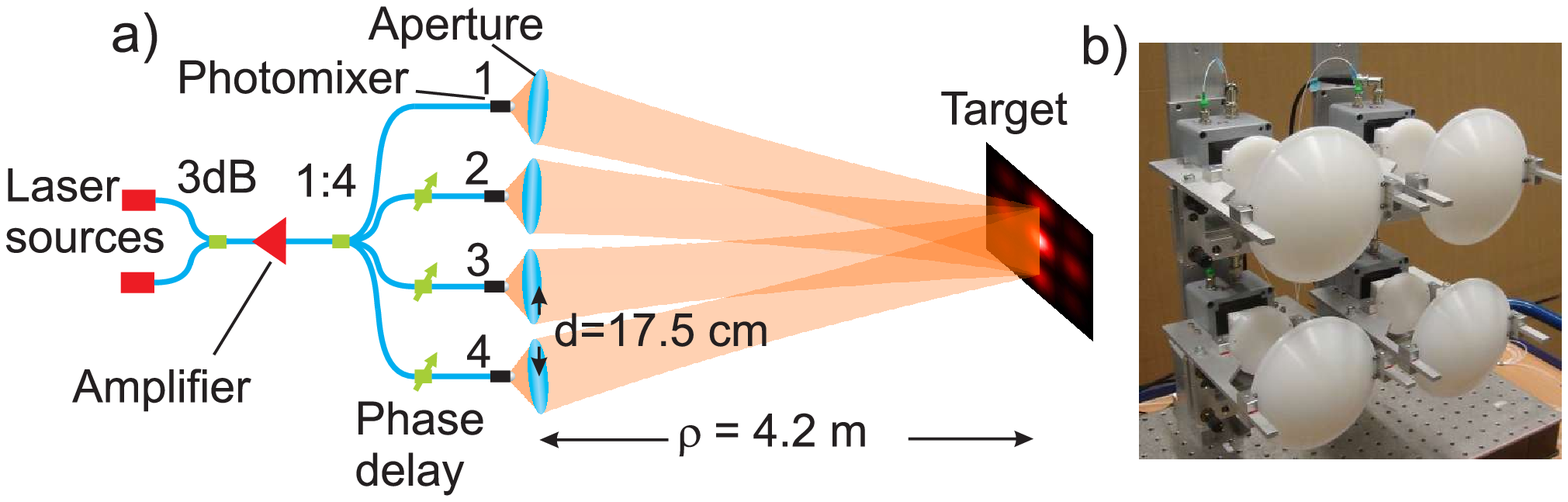}{10 cm}{h}{a) Schematic setup of the free standing photomixer array. b) Square 2$\times$2 emitter array with polyethylene lenses.%[** Das interference pattern zeigt 2x2 array, das schema aber ein 1x4 array, evtl. anderes bild nehmen)**]
}{Fig2}

A small phase offset between the emitters allows for steering the spot without mechanical tilting, similar to the on-chip array. Here, the phase offset is simply achieved by delaying the photomixing laser signal for the individual emitters by a fraction of the THz wavelength. Furthermore, the mutual coherence of $N$ in-phase emitters, each with its own focusing optics, generates an intense central spot in the target plane. The peak intensity of an $N$ element array is $N^2$ higher than that generated by a single device \cite{SPIE08}. At the same time, the beam waist is reduced by $w_{FWHM}^{array}=w_{FWHM}^{single}/(A\cdot N)$, where the parameter $A$ takes the spacing and aperture size of the sources into account. For a linear array with equidistant emitters spaced by a distance, $d$, as illustrated in \Figref{Fig2} a), and focused to a target at a distance, $\rho$, the interference of the individual beams generates an intense peak with a FWHM (full width half maximum) along the emitter axis of

\begin{equation}
w_{FWHM}^{array} \approx 0.95\frac{\lambda_{THz} \rho}{d}\cdot\frac{1}{N}.\label{eqnScaling}
\end{equation}

%The factor 0.95 relates the spacing of the first interference minima to the the FWHM of the central peak. 
The ideal diffraction-limited FWHM beam diameter of a single emitter in the target plane is

\begin{equation}
w_{FWHM}^{single} \approx 2\sqrt{ln2}\frac{\rho \lambda_{THz}}{\pi \w_A}.\label{eqnspotsize}
\end{equation} 

which is calculated by convoluting the emitted Gaussian beam with the circular aperture of the focusing lens (radius $w_A$). Dividing  \eqnref{eqnspotsize} by \eqnref{eqnScaling} yields for the parameter $A\approx0.56\ d/w_A$. $A$ is only dependent on the ratio of the source separation and the aperture diameter ($w_A<d$).

The photomixers used for the free space array are telecom-wavelength compatible n-i-pn-i-p superlattice photomixers \cite{MeinReview,APL1}. They are operated with fiber-coupled 1550 nm lasers that are amplified by an erbium-doped fiber amplifier. The amplified light is split up and distributed in the array in an all-polarization maintaining fiber setup. Care is taken to ensure the same optical path length of all devices. The photomixers are equipped with broadband logarithmic-periodic antennas with an operation range from 60 GHz to 2 THz and mounted on a hyperhemispherical silicon lens with a diameter of 10 mm. The generated THz beams are pre-shaped by a pair of aspherical polyethylene lenses (aperture diameter $2w_A=12$ cm of the focusing lens) in order to collimate the beam. The emitters are spaced by $d=$17.5 cm in a 4$\times$1 or a 2$\times$2 array. The individual THz beams are combined and interfered at a distance of $\rho=$ 4.2 m. We first examine the scaling law of \eqnref{eqnScaling} using a linear array with up to 4 sources. The recorded cross sections of the beam patterns in the target plane along the array axis for 1-4 emitters are shown in \Figref{Fig3V2} a) for a frequency of 0.23 THz. The measured peak width for these interference patterns are shown in table \ref{beamred}. The expected values from eq. (\ref{eqnScaling}), (\ref{eqnspotsize}) are also shown for comparison. The measured scaling factor of $A=1.5$ is in very good agreement with the expected value of 1.6.

%Hier Bild 3 in Finaler two-column Version
\bild{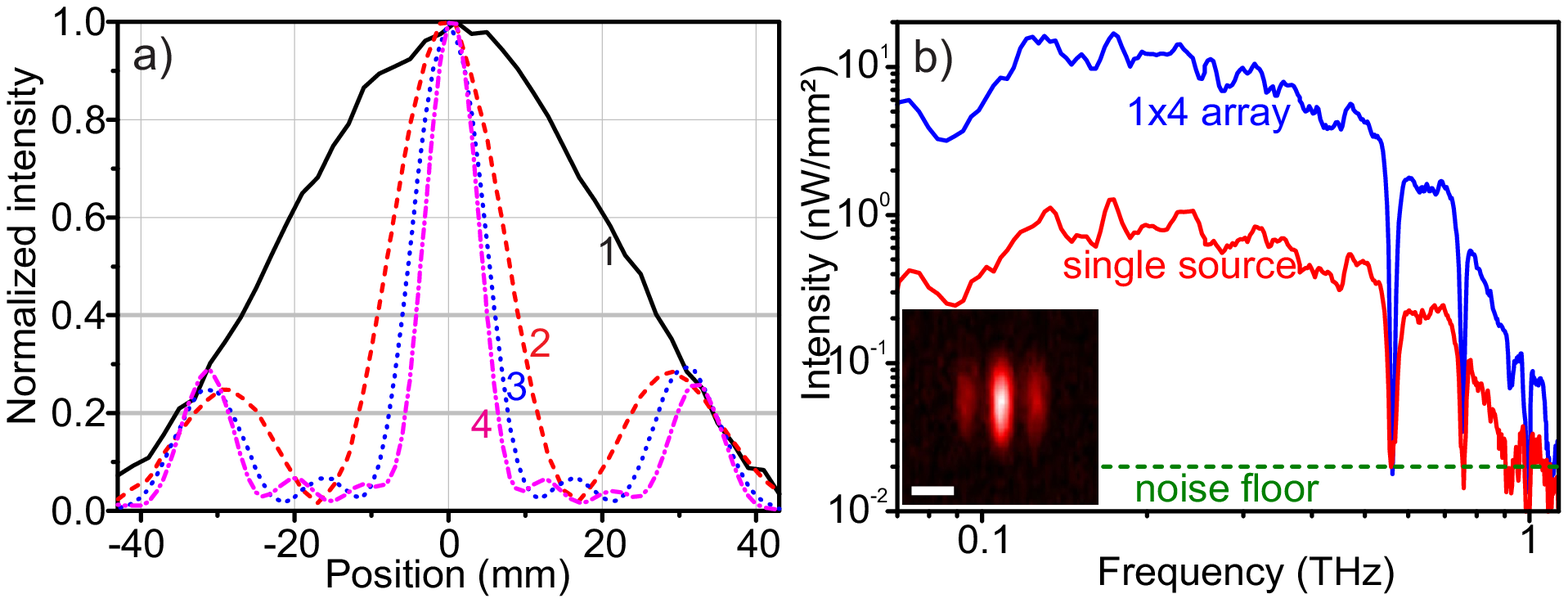}{13 cm}{h!}{a) Measured cross section (black dots) for arrays consisting of 1-4 elements ($d$=17.5 cm) in the linear configuration at a distance of 4.2 m. The cross sections were recorded along the array axis.  b) Transmitted power vs. frequency for the 4$\times$1 array and a single source in operation. A spectral range of 60 GHz to 1.1 THz is covered by the array. Water absorption lines are clearly resolved, particularly at high frequencies where the signal to noise ratio of the single sources is insufficient. The inset shows the interference pattern at 0.8 THz, the white scale bar is 10 mm.}{Fig3V2}

The data were obtained using a Golay cell detector with an aperture diameter of 6 mm. No focusing optics was used. The measured peaks are a convolution of the Golay cell aperture and the THz beam. The FWHM is therefore larger than prediced by theory. For 4 sources, the measured beam waist along the array axis is only 7.9 mm. This is well below the diffraction limit of a single device that features a FWHM beam waist of 47 mm. The same effect was verified at 0.81 THz. A slit aperture (width 2.25 mm) was used to scan the peak. A FWHM of 2.7 mm was recorded with 4 sources, again in the range of the slit size. The theoretical beam waist is 2 mm at this frequency.

The position of the central interference peak can be shifted within the envelope of the beam pattern of a single emitter by adjusting the relative phase of the individual photomixers. The strength of the zeroth order interference peak follows the envelope of the single source, resulting in reduced intensity for large shifts. Furthermore, higher interference orders (side lobes) gain strength once they approach the maximum of the single source beam pattern. The scan range should therefore be restricted to
$\sim \pm w_{FWHM}^{single}/2$ \cite{SPIESeb}.

\begin{table}
\begin{tabular}{|l|l|l|l|}\hline
\# of emitters &$w_{exp}$ (mm)&$w_{theo}$ (mm)\\\hline\hline
1&47&48\\\hline
2&15&14\\\hline
3&10&9.4\\\hline
4&7.9&7.0\\\hline
\end{tabular}
\caption{Measured ($w_{exp}$) and calculated ($w_{theo}$) peak width (FWHM) for different numbers of emitters in operation at 0.23 THz.}
	\label{beamred}
\end{table}

\Figref{Fig3V2} b) shows a frequency sweep of the 4$\times$1 array and of a single emitter measured with a Golay cell at 4.2 m distance in the laboratory without any additional focusing optics. The increase in intensity of almost a factor of $4^2=16$ persists over the whole frequency range. Water absorption lines are clearly visible.

%\bild{Fig3.eps}{8.5 cm}

\Figref{Fig4} a) shows the beam pattern of the 2$\times$2 array at 0.4 THz. \Figref{Fig4} b) and c) show the cross sections of the interference pattern at 0.32 THz. The intensity of the zeroth order (central peak) is almost $4^2=16$ times higher than that of an individual photomixer. The beam waist of the central peak of 11 mm is by a factor of 3.7 smaller than that of a single device (41 mm). 
Due to the small number of array elements, the side lobes are still very pronounced. They can be reduced by using more array elements. The side lobes along the array axis were suppressed by a factor of 2.4. The agreement between the theoretical expectation and the experiment is excellent.

\bild{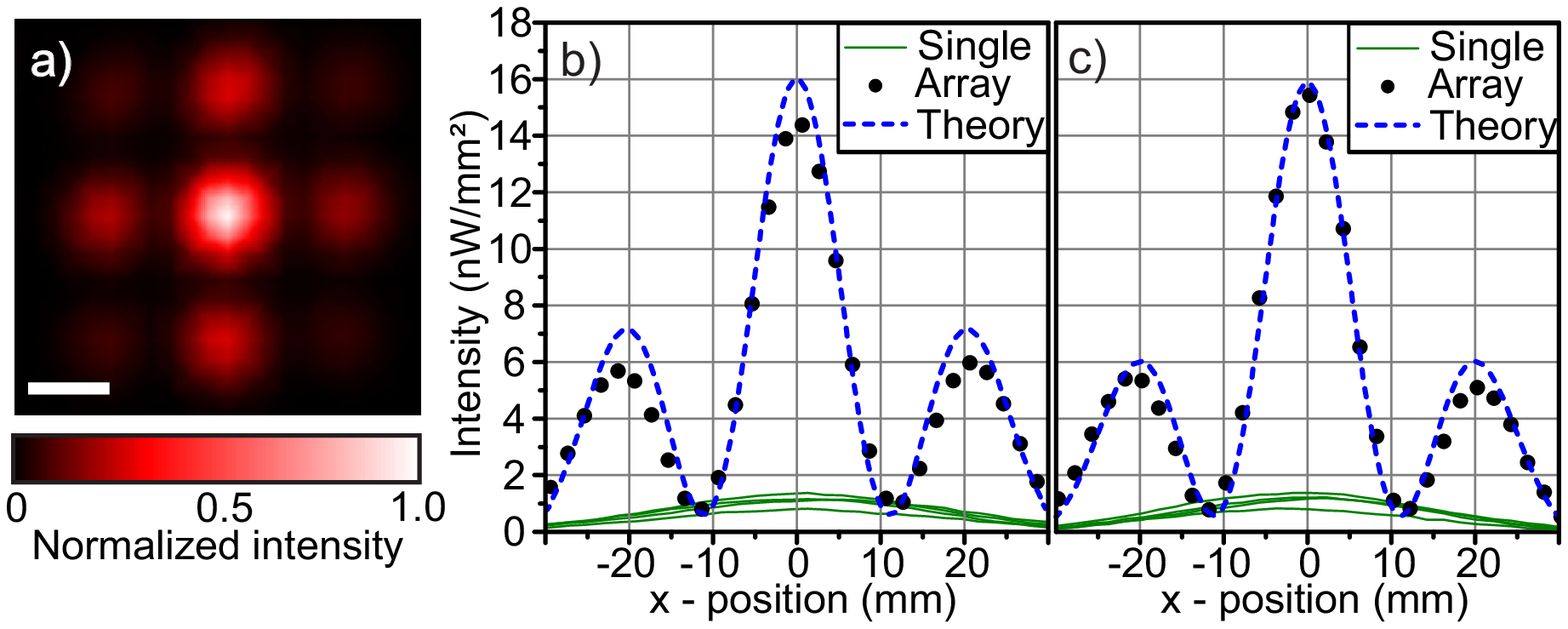}{11 cm}{h!}{a) Beam pattern of the 2$\times$2 array at 0.4 THz at a distance of 4.2 m, measured with the 6 mm Golay cell window without any additional focusing optics. The white scale bar is 10 mm. b) horizontal and c) vertical cross sections of the individual sources (green) and of the interference pattern of the 2$\times$2 array (dots) at 0.32 THz. The beam waist of the central spot is already comparable to the size of the Golay cell window of 6 mm. %The intensity of the central peak is therefore not well resolved, leading to just $\sim 14$ times higher intensity than that of a single device (green) instead of an improvement of $4^2=16$. 
The blue graph depicts the theoretical expectation according to ref. \cite{SPIE08}, taking the finite size of the Golay cell window into account.}{Fig4}

In conclusion, we characterized on-chip arrays and free space arrays of THz photomixers. On-chip arrays are of small size and allow for integratability at the cost of bandwidth due to cross talk. Free space arrays offer large bandwidth and tunability, small spot size ($\sim N^{-1}$) at stand-off distances due to interference effects and high focal intensity ($\sim N^2$) while using relatively small optics. The diameter of the interference peak for a linear array  at a distance of 4.2 m was more than 5 times smaller than that of a single emitter using 12 cm optics. Both array concepts allow for steering the THz beam by controlling the phases of the optical beams illuminating the array elements. This was demonstrated with the on-chip array by implementing a tilted phase front of the heterodyned laser beams. Fiber delay stages could be used for the fiber-coupled free space array for beam scanning. The advantages of both free space and chip-scale arrays could be combined by implementing broadband antennas in an on-chip array. The individual photomixers could be fiber-coupled, providing a high power, tunable, and miniaturized cw THz source. 

%[Quique zitieren; Nochmal?]

%Note: For reviewing: Journals WITH titles; For publication: Titles will be removed=> shorter!
%Et al ab 3 Authoren!!!!
%\bibliographystyle{osajnl}
%\bibliography{THz-generalMar2013}

\end{document}